\documentclass[final, 5p,times]{elsarticle}
\usepackage{graphicx} 
\graphicspath{{images/}} 
\usepackage{url}

\usepackage{subcaption}

\usepackage{amssymb,amsmath}

\usepackage{booktabs}
\usepackage{multirow}
\usepackage[colorinlistoftodos]{todonotes}
\usepackage{comment}
\usepackage{color, soul}
\usepackage{microtype}
\soulregister\cite7
\soulregister\ref7
\soulregister\pageref7

\makeatletter
\def\ps@pprintTitle{%
	\let\@oddhead\@empty
	\let\@evenhead\@empty
	\let\@oddfoot\@empty
	\let\@evenfoot\@empty}
\makeatother

\begin{document}

\begin{frontmatter}

\title{Secure Federated Data Distillation}

\author[address1]{Marco Arazzi}
\ead{marco.arazzi01@universitadipavia.it}
\affiliation[address1]{
organization={Department of Electrical, Computer and Biomedical Engineering, University of Pavia}, 
addressline={A. Ferrata, 5},
city={Pavia},
postcode={27100},
state={PV},
country={Italy}
}

\author[address1]{Mert Cihangiroglu}
\ead{mert.cihangiroglu01@universitadipavia.it}

\author[address2]{Serena Nicolazzo\corref{correspondingauthor}}
\ead{serena.nicolazzo@unimi.it}
\affiliation[address2]{
organization={Department of Computer Science, University of Milan}, 
addressline={G. Celoria, 20},
city={Milan},
postcode={20133},
state={MI},
country={Italy}
}

\author[address1]{Antonino Nocera}
\cortext[correspondingauthor]{Corresponding author}
\ead{antonino.nocera@unipv.it}

\begin{abstract}
Dataset Distillation (DD) is a powerful technique for reducing large datasets into compact, representative synthetic datasets, accelerating Machine Learning training.
However, traditional DD methods operate in a centralized manner, which poses significant privacy threats and reduces its applicability. To mitigate these risks, we propose a Secure Federated Data Distillation (SFDD) framework to decentralize the distillation process while preserving privacy.
Unlike existing Federated Distillation techniques that focus on training global models with distilled knowledge, our approach aims to produce a distilled dataset without exposing local contributions. We leverage the gradient-matching-based distillation method, adapting it for a distributed setting where clients contribute to the distillation process without sharing raw data. The central aggregator iteratively refines a synthetic dataset by integrating client-side updates while ensuring data confidentiality. To make our approach resilient to inference attacks perpetrated by the server that could exploit gradient updates to reconstruct private data, we create an optimized Local Differential Privacy approach, called LDPO-RLD (Label Differential Privacy Obfuscation via Randomized Linear Dispersion). Furthermore, we assess the framework's resilience against malicious clients executing backdoor attacks (such as Doorping) and demonstrate robustness under the assumption of a sufficient number of participating clients. Our experimental results demonstrate the effectiveness of SFDD and that the proposed defense concretely mitigates the identified vulnerabilities, with minimal impact on the performance of the distilled dataset. By addressing the interplay between privacy and federation in dataset distillation, this work advances the field of privacy-preserving Machine Learning making our SFDD framework a viable solution for sensitive data-sharing applications.

\end{abstract}




\end{frontmatter}


\section{Introduction}
Dataset Distillation (DD, hereafter) is defined as a set of approaches designed to generate compact yet highly informative data summaries to capture the essential knowledge of a given dataset. These distilled representations are optimized to function as efficient substitutes for the original dataset, enabling accurate and resource-efficient applications such as model training, inference, and architecture search \cite{wang2018dataset}.
DD has attracted much attention from the deep learning community because it addresses the problem of handling unlimited data growth with limited computing power. 

Typically, DD methods operate within a centralized and static framework, where the entire dataset is accessible at a single location \cite{wang2018dataset,cazenavette2022dataset,zhao2020dataset}. 
In particular, a straightforward approach to constructing a synthetic dataset implies that each data owner shares its data with a central server so DD can happen. Despite the numerous benefits, concentrating information in a single point of aggregation may lead to privacy leakages. Indeed, the central server may be honest but curious (or even malicious) and might take advantage of all the shared sensitive information.
One specific application domain, in which this aspect may represent a critical issue is the sharing of medical datasets to establish the cross-hospital flow of medical information and improving the quality of medical services (for instance to construct high-accuracy computer-aided diagnosis systems) \cite{kumar2021integration,weitzman2010sharing}. In this context, the different hospitals should share the medical information of all their patients with an external entity that is responsible for distilling the data before starting the specific analysis. In most cases, patients are reluctant to share such highly sensitive information, making DD approaches impractical \cite{aouedi2022handling}.

Borrowing some ideas from the new paradigm of Federated Learning (FD, hereafter), which distributes the learning process across multiple entities to enhance privacy, our work proposes a novel framework for Secure and Federated Data Distillation (SFDD). This approach seeks to enable efficient DD while preserving data privacy by decentralizing the distillation process across multiple participants. This ensures that raw data remains local while only distilled knowledge is shared.
Observe that our proposal takes a different perspective 
from the existing Federated Distillation (FD) schemes \cite{zhou2020distilled,cazenavette2022dataset,song2023federated}. With the objective of improving FL performance and privacy guarantees, FD distills knowledge from multiple local models independently and transfers only compact, distilled information to the server that trains a global model using this distilled data. By contrast, the aim of our approach is to collaboratively distill data to produce a common distilled dataset without sharing local information. 
To do so, we start considering the method proposed by Zhao et al. \cite{zhao2020dataset} for learning a synthetic set such that a deep network trained on it would preserve similar performance to that obtained when trained on the original large dataset. The synthetic data can later be used to train a network from scratch in a small fraction of the original computational load. We adopt this method and we make it distributed among different data owners (or clients). Firstly, a central server (or aggregator), that is in charge of aggregating the different contributions, produces and shares a random synthetic set of data. Then, the clients create local distillation contributions and at each step, they optimize the synthetic set of data and return to the server an enhanced version for aggregation till convergence is obtained.

However, in the pure FL context, some work \cite{zhu2019deep} have demonstrated that it is possible to obtain private training data from publicly shared gradients. Hence an honest but curious server may take advantage of the obtained updates sent by the clients to reconstruct the original batch of data and leak the privacy of data owners. For this reason, we first assess the vulnerability of our approach to this threat. We then enhance its security by implementing an improved Local Differential Privacy (LocalDP) strategy, called LDPO-RLD (LabelDP Obfuscation via Randomized Linear Dispersion). Adding this defense to our SFDD framework allows the clients to obfuscate the point-to-point correlation between distilled images and real ones. Experiments demonstrate that including LDPO-RLD in SFDD is not only an effective defense against deep leakage attacks but also outperforms the standard LocalDP in distillation performance. 

Additionally, we test our solution also in the presence of malicious clients. Indeed, a recent work introduced a new backdoor attack method, called Doorping, which attacks during the dataset distillation process rather than after the model training \cite{liu2023backdoor}. Our experimental campaign demonstrates that under the assumption of a sufficient number of clients, our framework is robust also to this new type of attack. 

The results obtained during our experimental campaign demonstrate that our Federated Data Distillation approach is secure against known threats.

In summary, the key contributions of our framework are the following.

\begin{itemize}
    \item We propose a strategy that allows diverse data owners to participate in a global and fully distributed Data Distillation process without sharing local data. DD is computed by merging all the contributions of the different clients, thus no raw data or detailed model parameters are exchanged, and privacy is preserved.
    \item Our framework is also secure against inference attacks. In fact, honest-but-curious servers cannot infer sensitive information about clients data by exploiting the gradient and update dynamics exchanged during the distillation process thanks to the presence of our LabelDP Obfuscation via Randomized Linear Dispersion (LDPO-RLD) defense strategy.
    \item We also demonstrate that, under the assumption of a sufficient number of clients, our framework is robust to client-side attacks, such as the Doorping attack.
\end{itemize}

The outline of this paper is as follows. In Section \ref{sec:related}, we examine the related papers present in the state-of-the-art. Section \ref{sec:background} is devoted to describing some basic concepts related to Federated Learning and Dataset Distillation useful for comprehending our proposal. Section \ref{sec:description} gives a general overview of our reference model and details the proposed framework. In Section \ref{sec:experiments}, we present the experiments carried out to test our approach and show its performance. Finally, Section \ref{sec:conclusion} examines intriguing leads as future work and draws our conclusions.

\section{Related Work}
\label{sec:related}

Dataset Distillation (DD, hereafter) \cite{wang2018dataset} has recently emerged as a novel paradigm to synthesize a significantly
smaller dataset from a large dataset, aiming to maintain the same training accuracy performance as if it was trained on the original large dataset. 
In this section, we will describe the proposals that combine FL with Distillation.
The work presented in \cite{li2019fedmd,zhu2021data,jeong2018communication,lin2020ensemble,afonin2021towards} leverage Knowledge Distillation (KD, for short) to transfer knowledge from local client models to a centralized FL server model to improve FL performance.
In particular, \cite{li2019fedmd} describes FedMD, a new FL framework that allows participants to independently design their models and maintain them private due to privacy and intellectual property concerns. To collaborate with each client, owning a private dataset and a uniquely designed model, FedMD relies on KD to translate its learned knowledge into a standard format.
The authors of \cite{zhu2021data} try to solve the classical FL issue of data heterogeneity (i.e., data from the real world are usually not independent and identically distributed) by proposing a data-free Knowledge Distillation approach for FL. This framework extracts the knowledge from users without depending on any external data through KD and then it regulates local model updating using this extracted knowledge.
Instead, the proposed framework called FD \cite{jeong2018communication} utilizes distillation to reduce FL communication costs. It synchronizes logits per label which are accumulated during the local training. The averaged
logits per label (over local steps and clients) are exploited as a distillation regularizer for the next round's local training. Moreover, Lin et al. \cite{lin2020ensemble} propose ensemble distillation for model fusion with the aim of training
the central classifier through unlabeled data on the outputs of the models from the clients. 
Finally, Afonin and Karimireddy \cite{afonin2021towards}
describe a theoretical framework, called Federated Kernel
ridge regression, which is a model-agnostic FL scheme allowing each agent to train
their model of choice on the combined dataset.
All the above works propose KD-based Federated Learning protocols with a purpose that is quite different from ours. Indeed, firstly, they deal with Knowledge Distillation instead of Data Distillation, secondly, we aim to design a new scheme to distribute DD on FL, but we do not aim to improve the FL protocol through Distillation.

Only a little research has explored Dataset Distillation approaches in FL. For instance, the strategies proposed in \cite{zhou2020distilled,song2023federated,cazenavette2022dataset} try to reduce the communication cost of FL while achieving comparable performance by exploiting DD. In these protocols, instead of sharing model updates, the clients distill
the local datasets independently in just one round, and then they send the synthetic data (e.g. images or sentences) to the central server that aggregates those decentralized distilled datasets. In these works, similarly to ours, DD is used instead of KD. In particular, Zhou et al. \cite{zhou2020distilled} propose a method combining DD and distributed one-shot learning. For every local update step, each synthetic data successively updates the network for one gradient descent step. Thus, synthetic data is closely related to one specific network weight, and the eavesdropper cannot reproduce the result with only leaked synthetic data.

As already said we employ a different perspective. Indeed in the above-cited approaches, DD has been applied to FL to develop a
privacy-preserving distributed model training scheme such that
multiple clients collaboratively learn a model without sharing
their private data. Instead of transmitting model updates as the standard way of FL, which may cause demanding communication costs they transmit the locally generated synthetic datasets proved for
privacy protection and essence information preservation. Instead, we aim to improve the Dataset Distillation framework by federating it.
To the best of our knowledge, our proposed scheme is the first of its kind.

\section{Background}
\label{sec:background}

In this section, we illustrate the main concepts that can be useful for a clear understanding of our approach. In particular, we focus on the description of Federated Learning (FL) and Knowledge Distillation (KD) mechanisms.

Table \ref{tab:SystemSymbols} summarizes the acronyms used in this paper.

\begin{table}
\centering
  \caption{Summary of the acronyms used in the paper}
  \resizebox{\columnwidth}{!}{
  	\begin{tabular}{ll}
\hline
    \textbf{Symbol} & \textbf{Description}\\
\hline
    DD & Dataset Distillation\\
    DL & Deep Learning\\
    FD & Federated Distillation\\
    FL & Federated Learning\\
    GM & Global Model \\
    IID & Independent and Identically Distributed\\
    KD & Knowledge Distillation\\
    LDPO-RLD & LabelDP Obfuscation via Randomized Linear Dispersion\\
    LocalDP & Local Differential Privacy\\
    LM & Local Model\\
    MD & Model Distillation\\
    ML & Machine Learning\\
    IPC & Image Per Class\\
    SFDD & Secure Federated Data Distillation\\
\hline
\end{tabular}
}
\label{tab:SystemSymbols}
\end{table}

\subsection{Federated Learning}

Federated Learning is designed to train an ML model in a decentralized manner across different devices holding local data samples. Keeping local data confidential without exchanging them with other participants or a central server allows privacy preservation; whereas sending only model updates reduces communication overhead and network traffic.

\begin{figure}
    \centering
    \includegraphics[width=0.8\columnwidth]{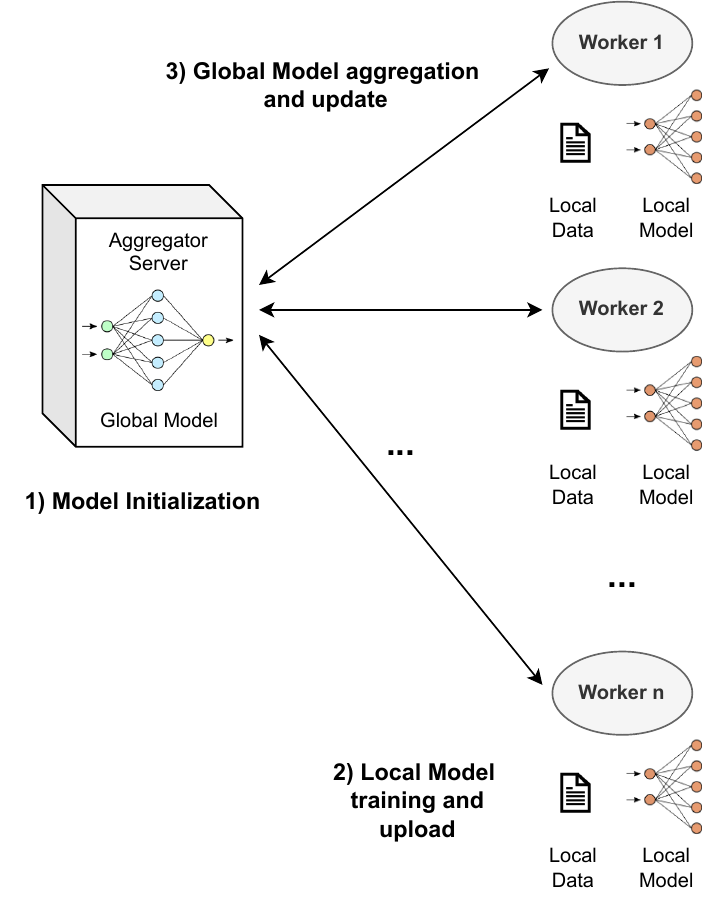}
    \caption{The Federated Learning workflow}
    \label{fig:FL}
\end{figure}

As visible in Figure \ref{fig:FL}, the participants of this protocol are mainly of two types:

\begin{itemize}
    \item the {\em worker} nodes, also called {\em clients}, that are $\mathcal{C}$ devices executing local training with their private data;
    \item an {\em aggregator} node, or {\em central server}, which is in charge of the coordination of the whole FL approach and aggregates the local updates. 
\end{itemize}

Hence, the main goal of FL is to train a Global Model (GM), say $\mathbf{w}$, by uploading the weights of Local Models (LMs) 
$\{\mathbf{w}^i|i \in \mathcal{C}\}$ to the central server. Equation \ref{eq:lossFunction} shows the loss function to be optimized:

\begin{equation}
    \min\limits_{\mathbf{w}} l(\mathbf{w}) = \sum_{i=1}^n{\frac{s_i}{\mathcal{C}}L_i(\mathbf{w}^i)}
    \label{eq:lossFunction}
\end{equation}

\noindent
where $L_i(\mathbf{w^i})= \frac{1}{s_i}\sum_{j \in I_i}{l_j(\mathbf{w}^i, x_i)}$ is the loss function, $s_i$ is the local data size of the {\em i}-th worker, and $I_i$ identifies the set of data indices
with $|I_i|=s_i$, and $x_j$ is a data point.

The basic FL workflow can be divided into three main phases \cite{zhang2021survey}. 
During the first stage, called {\em Model initialization}, the server {\em (i)} initializes the necessary parameters for the GM $\mathbf{w}$; and {\em (ii)} select the workers for the FL process. 
The second phase consists of the {\em LMs training and uploading}. The clients download the current GM and perform local training on their private data during this stage. After that, each client computes the model parameter updates and sends them to the server. The regional training involves more than one iteration of back-propagation, gradient descent, or other optimization methods to improve the LM's performance. In particular, for each iteration, the different clients update the GM with their datasets: $\mathbf{w}^i_t \leftarrow \mathbf{w}^i_t - \eta \frac{\partial L(\mathbf{w}_t,b)}{\partial \mathbf{w}^i_t}$, where $\eta$ specifies the learning rate and $b$ is the local batch.
Finally, the {\em GM aggregation and update} phase is performed. In this step, the server collects and aggregates the model parameter updates from all the workers, $\{\mathbf{w}^i|i \in \mathcal{C}\}$. The aggregator can employ various methods like averaging, weighted averaging, or secure multi-party computation (SMC) to incorporate the received updates from each client.

\begin{figure}
\centering
\begin{subfigure}{0.3\textwidth}
    \includegraphics[width=\textwidth]{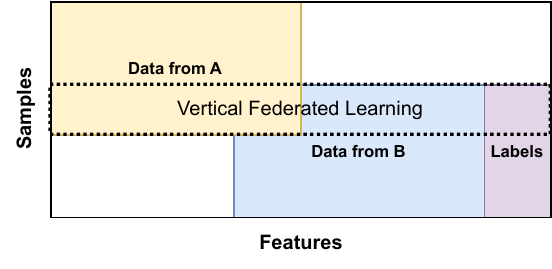}
    \caption{VFL}
    \label{fig:vfl}
\end{subfigure}
\begin{subfigure}{0.3\textwidth}
    \includegraphics[width=\textwidth]{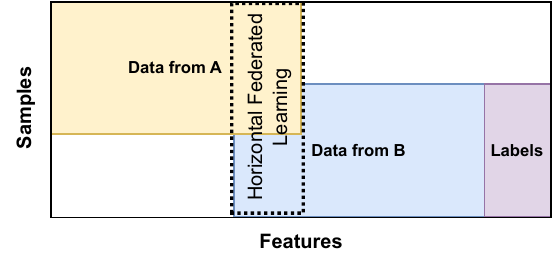}
    \caption{HFL}
    \label{fig:hfl}
\end{subfigure}
\begin{subfigure}{0.3\textwidth}
    \includegraphics[width=\textwidth]{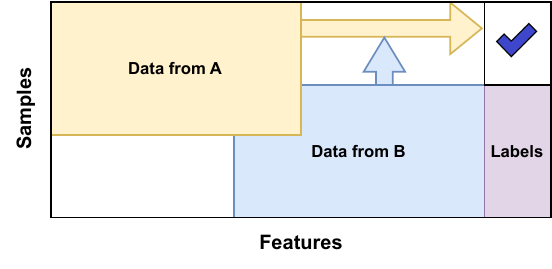}
    \caption{FTL}
    \label{fig:ftl}
\end{subfigure}
\caption{The three categories of FL divided for feature and sample spaces}\label{fig:HorizontaleVerticalFTL}
\end{figure}

As visible in Figure \ref{fig:HorizontaleVerticalFTL}, FL can assume the following three configurations according to the different data partition strategies considered \cite{yang2019federated}:

\begin{itemize}
    \item \textbf{Vertical Federated Learning} (VFL) in the case in which the datasets share overlapping data samples but differ in the feature space (see Figure \ref{fig:vfl}). This scheme can be applied if two different organizations (i.e., an Internet service provider and an online TV streaming provider) have data about the same group of people with different features and want to collaboratively train an ML model while keeping their data private.
    \item \textbf{Horizontal Federated Learning} (HFL) that is used for cases in which each device contains a dataset with the same feature space but with different sample instances. For instance, think of two branches of the same insurance company that hold the same type of data about different clients (see Figure \ref{fig:hfl}).
    \item \textbf{Federated Transfer Learning} (FTL) borrows some characteristics from both VFL and HFL and is suitable for scenarios in which there is little overlapping in both data samples and features as visible in Figure \ref{fig:ftl}. A good example is the case in which a bank wants to train its ML model by cooperating with an insurance company that shares part of the client and part of the features.
\end{itemize}

Even if FL has been designed to achieve data confidentiality it has been demonstrated that it is still prone to possible attacks targeting data privacy that any participants of the scheme can perpetrate \cite{lyu2022privacy}. The most common attacks in this context are the following:

\begin{itemize}
    \item \textbf{Inference attacks} that aim at inferring the sensitive information about individual data points (attribute inference) or participants (membership inference) in the training dataset by analyzing the behavior or outputs of the federated model \cite{nasr2019comprehensive,arazzi2025defense,arazzi2023blindsage}.
    \item \textbf{Poisoning attacks} can be divided into data or model poisoning attacks. The first category involves adversaries that try to poison the training data in a certain number of devices participating in the learning process to compromise the GM accuracy. The adversary can inject poisoned data {\em (i)} directly into the targeted device or {\em (ii)} through other devices \cite{sun2021data, arazzi2023turning}. In a model poisoning attack the adversary tries to poison the LMs instead of the local data to introduce errors in the GM.
    \item \textbf{Backdoor attacks} through which an adversary can mislabel certain tasks without affecting the accuracy of the GM. This kind of attack manipulates a subset of training data by injecting adversarial triggers such that the models trained on the tampered dataset will make arbitrarily (targeted) incorrect predictions on the test set with the same trigger embedded \cite{gu2019badnets, arazzi2024let}.
\end{itemize}

\subsection{Dataset Distillation}
\label{sub:centralizedDistillation}
In general, in the context of ML, Distillation (known as Model Distillation) is a methodology to transfer knowledge from a larger, more complex model (called ``teacher'') to a smaller, simpler model (known as ``student'') to improve model performance or deploy the model on resource-constrained devices, such as Internet of Things (IoT) devices.

\begin{figure}
    \centering
    \includegraphics[width=0.9\columnwidth]{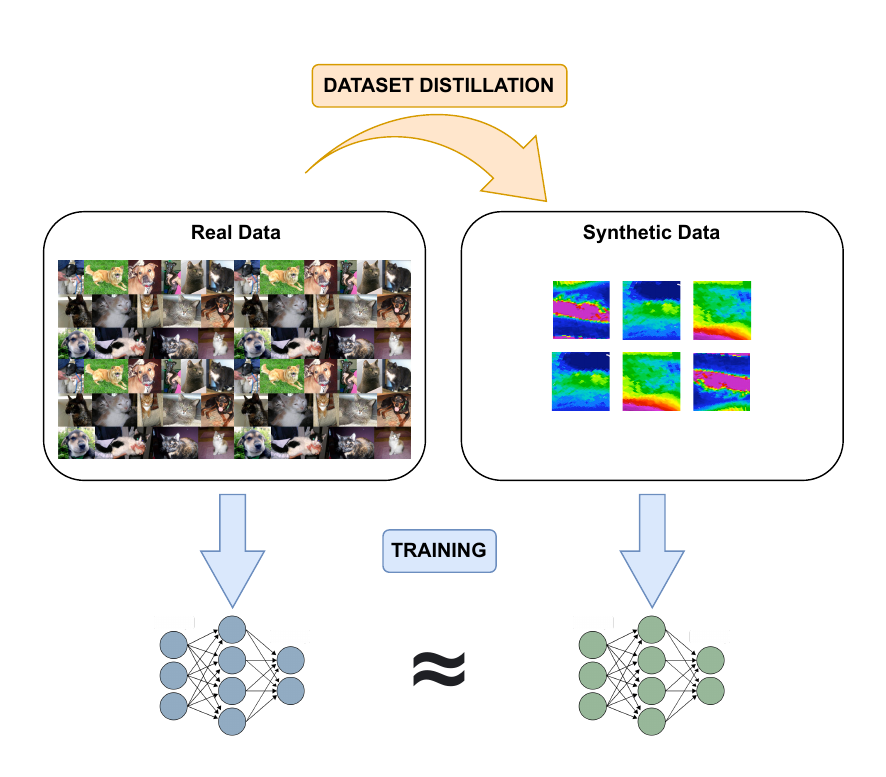}
    \caption{Dataset Distillation (DD) scheme}
    \label{fig:DD}
\end{figure}

An alternative concept proposed by \cite{wang2018dataset} is called Dataset Distillation (DD) and consists of the summarization of real data in a few highly informative and synthetic data points in such a way that models trained on the last dataset achieve comparable generalization performance to those trained on the real data. 
has been created to address this problem of large data volume \cite{lei2023comprehensive}. Figure \ref{fig:DD} illustrates the Dataset Distillation scheme, showing that the models trained on the large original dataset and small synthetic dataset demonstrate comparable performance
on the test set.

To formally define DD, we start with some preliminary definitions, namely a target dataset:

$$\mathcal{T} = \{(x_i, y_i)\}^m_{i=1}$$

\noindent
where $x_i \in \mathbb{R}^d$, $d$ is the dimension of the input data, $y_i$ is the $i-th$ label, and $(x_i, y_i)$, with $1 \leq i \leq m$ are independent
and identically distributed (i.i.d.) random variables drawn from the data generating distribution $\mathcal{D}$.
The goal of DD is to extract the knowledge of $\mathcal{T}$ into a small synthetic dataset called:

$$\mathcal{S} = \{(s_j, y_j)\}^n_{j=1}$$

\noindent
where $n<<m$ and the model trained on the small distilled dataset $\mathcal{S}$ can achieve a generalization performance that can be approximated to the one of the original dataset $\mathcal{T}$:

$$\mathbb{E}_{\substack{(x,y)\sim\mathcal{D}\\ \Theta^{(0)}\sim \mathcal{P}}} [\ell( f_{alg(\mathcal{T})}(x), y)] \simeq \mathbb{E}_{\substack{(x,y)\sim\mathcal{D}\\ \Theta^{(0)}\sim \mathcal{P}}} [\ell( f_{alg(\mathcal{S})}(x), y)]  $$

\noindent
where 

\begin{itemize}
    \item $\Theta^{(0)}$ is the initialized network parameter;
    \item $f_{alg(\mathcal{T})}$ is the model $f$ trained on the original dataset $\mathcal{T}$;
    \item $f_{alg(\mathcal{S})}$ is the model $f$ trained on the synthetic dataset $\mathcal{S}$;
    \item $f_{alg(\mathcal{\bullet})}(x)$ is the prediction or output of $f_{alg(\mathcal{\bullet})}$ at $x$;
    \item $\ell( f_{alg(\mathcal{\bullet})}(x), y)$ is the loss between the prediction $f_{alg(\mathcal{\bullet})}(x)$ and ground truth $y$;
    \item $\mathbb{E}_{\substack{(x,y)\sim\mathcal{D}\\ \Theta^{(0)}\sim \mathcal{P}}} [\ell( f_{alg(\mathcal{\bullet})}(x), y)]$ is the empirical risk and refers to the average loss or error of a model on a training dataset.
\end{itemize}

\section{Description of Our Approach}
\label{sec:description}

In this section, we provide a detailed description of our proposal.
In particular, in Section \ref{sub:generalOverview}, we give an overview of our approach and its underlying model, formally defining the {\em(i)} the involved parties and {\em(ii)} the Secure Federated Data Distillation (SFDD) process. After that, in Section \ref{sub:dataLeakage}, we describe a possible vulnerability of our scheme to inference attack and the modification we applied to make it robust to this particular kind of attack.

\subsection{General Overview}
\label{sub:generalOverview}

This section describes our SFDD architecture for decentralized and secure Data Distillation. 
We start by defining the fundamental components of our strategies, in particular, the involved parties are:

\begin{itemize}
    \item a {\em Central Unit} $C$ that acts as an aggregator node and is in charge of initializing the global synthetic set of data $\mathcal{S}^g$, randomly;
    \item a set of {\em Workers} or clients $\mathcal{W} = \{w_1, \dots, w_z\}$ of size $z$, which execute the Federated Distillation algorithm.
\end{itemize}

In our configuration, each worker $w_i$ in $\mathcal{W}$ holds a private data set $\mathcal{T}^{w_i} = \{(x^z_j, y^z_j)\}^m_{j=1}$ to be distilled. All the private datasets of the network are independent and identically distributed (IID) and do not overlap. The set composed of all the $\mathcal{T}^{w_i}$ for each worker $w_i$ can be formally defined as follows:

$$\mathcal{T}^{\mathcal{W}} = \{\mathcal{T}^{w_1}, \dots, \mathcal{T}^{w_z}\}; \ \mathcal{T}^{w_i} = \{(x^z_j, y^z_j)\}^m_{j=1}$$

\noindent
where $m$ is the number of data points for a set.

Our adopted technique relies on gradient matching and follows the main step described in \cite{zhao2020dataset}.
According to this strategy, each worker $w_k \in \mathcal{W}$ holds a private model $\mathcal{M}^{w_k}$ used to transfer knowledge from its private set of data $\mathcal{T}^{w_k}$ to the global synthetic set $\mathcal{S}^g$. The set of private models can be referred to as:

$$\mathcal{M}^{\mathcal{W}} = \{\mathcal{M}^{w_1}, \dots, \mathcal{M}^{w_z}\}$$

In the following, we detail all the steps of our Secure Federated Data Distillation approach. In brief, the steps are:
\begin{enumerate}
    \item Data Initialization;
    \item Local Data Distillation Update;
    \item Global Data Distillation Aggregation.
\end{enumerate}

The first step of our framework, called {\em Data Initialization}, is performed by the central unit $C$ that starts the process by randomly initializing the synthetic set of data $\mathcal{S}^g$. Specifically, it initializes a given number $ipc$ of data (representing the number of images per class) for each class of the original dataset as follows:

$$\mathcal{S}^g = \{\mathcal{S}^g_0, \dots, \mathcal{S}^g_c, \dots, \mathcal{S}^g_{n_c}\}; \ \mathcal{S}^g_c = \{(s_{i_c}, y_c)\}^{ipc}_{i_c=1}$$

\noindent
where $y_c$ is the label assigned to class $c$ and $n_c$ is the total number of classes.
The synthetic dataset $\mathcal{S}^g$ is distributed to all the workers in $\mathcal{W}$.

Once each worker receives $\mathcal{S}^g$, the second step of the framework of {\em Local Data Distillation Update} takes place and the local distillation phase of the process carried out by the workers starts in a parallel and independent way.
In particular, each worker $w_k$ distills the data contained in its private set $\mathcal{T}^{w_k}$ using its private model $\mathcal{M}^{w_k}$.

The distillation process is conducted separately for each class and it generates the gradients of the model $\mathcal{M}^{w_k}$ using two contributions, namely {\em(i)} the synthetic data $\nabla \mathcal{M}^{w_k}_{S^g_c}$ and {\em(ii)} a batch of real data $\nabla \mathcal{M}^{w_k}_{\mathcal{T}^{w_k}_c}$ (where $\mathcal{T}^{w_k}_c$ is a subset of $\mathcal{T}^{w_k}$ in which $y^z_i=y_c$).

Then, the obtained partial gradients are used to compute the matching loss using the matching loss function $ML$ of the considered category.

The {\em Local Data Distillation Update} step is repeated for all the classes in the dataset and its outcome is a total loss that is back-propagated onto the global synthetic dataset $\mathcal{S}^g$. More formally, this step can be formulated as follows:

\begin{gather}
    \nabla\mathcal{M}^{w_k}_{S^g_c} \leftarrow CE(\mathcal{M}^{w_k}(\mathcal{S}^g_c), y_c); \nabla \mathcal{M}^{w_k}_{\mathcal{T}^{w_k}_c} \leftarrow CE(\mathcal{M}^{w_k}(\mathcal{T}^{w_k}_c), y_c) \\
    \mathcal{S}^g_{w_k} = \mathcal{S}^g \leftarrow \sum^{n_c}_{c=1} ML(\nabla \mathcal{M}^{w_k}_{S^g_c}, \nabla \mathcal{M}^{w_k}_{\mathcal{T}^{w_k}_c}).
\end{gather}

\noindent 
\\
where $CE(\cdot)$ is the cross-entropy loss used to compute the gradients on $\mathcal{M}^{w_k}$ and $\mathcal{S}^g_{w_k}$ is the locally updated version of $\mathcal{S}^g$ obtained by the worker $w_k$.
All the $\mathcal{S}^g_{w_k}$ data are then used to train the local model $\mathcal{M}^{w_k}$ before the next iteration of the Federated Distillation. 

The last step of the framework is the {\em Global Data Distillation Aggregation}, in which the obtained updates on $\mathcal{S}^g_{w_k}$ are sent back to the central unit $C$ to be aggregated using the FedAvg strategy and sending back to the workers the new global $\mathcal{S}^g$.

\begin{figure*}[!ht]
    \centering
    \includegraphics[width=0.9\textwidth]{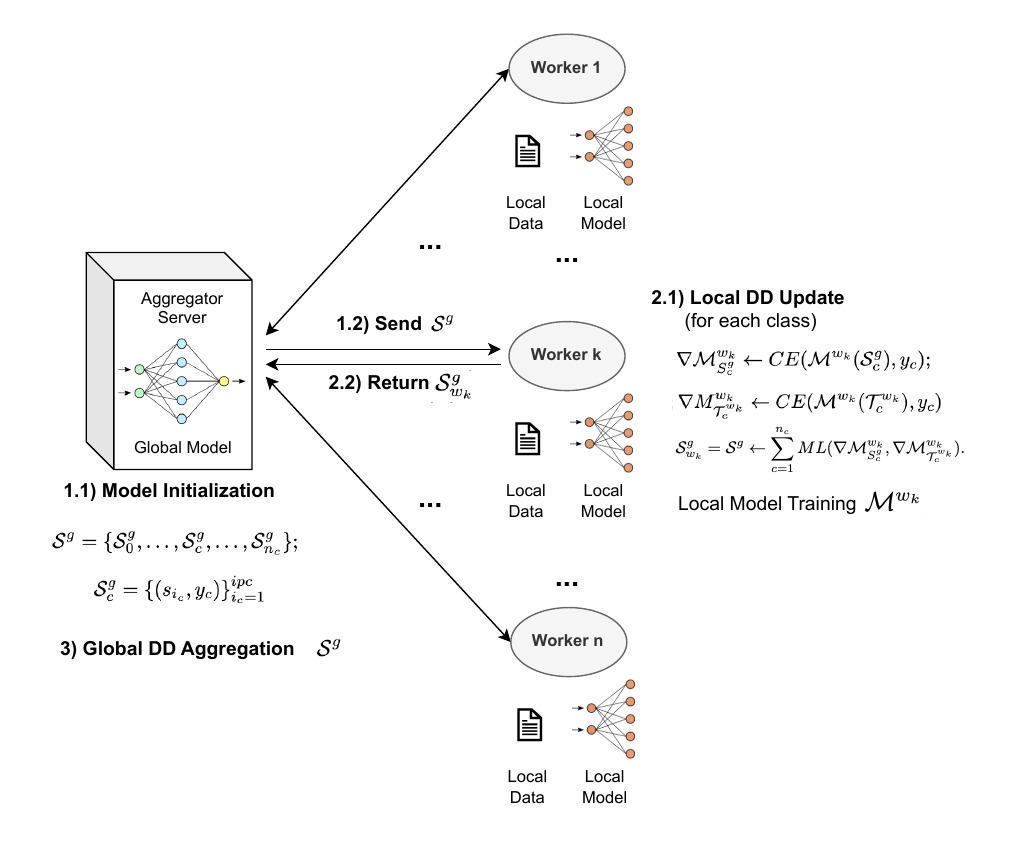}
    \caption{Secure Federated Data Distillation (SFDD) Architecture}
    \label{fig:architecture}
\end{figure*}

The last two steps of the process are repeated until the model's accuracy on the distilled dataset converges.
Figure \ref{fig:architecture} shows all the steps performed by our framework.

\subsection{Defence against Data Leakage Attack}
\label{sub:dataLeakage}

Like traditional Federated Learning strategies, as tested experimentally in Section \ref{sec:experiments}, our approach is vulnerable to data leak attacks perpetrated by an honest but curious server \cite{zhu2019deep}. This attack is carried out to obtain the private training data from the publicly shared
gradients. To perform it, a server takes advantage of the updates returned by the different workers to reverse the distillation process and obtain the original batch of data $\mathcal{T}^{w_k}_c$ used in the current epoch by the worker $w_k$.

In particular, in this context, the server can randomly initialize a tensor $\mathcal{LB}$ of the same shape as the one used by the workers to distill the knowledge ($batch-size \ X \ n_c$). The tensor is defined as follows:

$$\mathcal{LB}^{w_k} = \{\mathcal{LB}^{w_k}_0, \dots, \mathcal{LB}^{w_k}_c, \dots, \mathcal{LB}^{w_k}_{n_c}\}$$

\noindent
where $\mathcal{LB}_c$ is the leaked batch used by $w_k$ to distill the synthetic data $\mathcal{S}^g_c$ for the corresponding class $c$.
To reverse the process, the server tries to replicate the distillation strategy by freezing the $\mathcal{S}^g$ at the previous step and calculating the updates $\nabla \mathcal{S}^{g'}_{w_k}$ emulating the workers but using $\mathcal{LB}^{w_k}_c$ instead of $\mathcal{T}^{w_k}_c$ for all the classes. In this way, the server tries to replicate and match the updates $\nabla \mathcal{S}^g_{w_k}$ on $\mathcal{S}^g$ returned by the worker $w_k$.
The process can be formalized as follows:

\begin{gather}
    \nabla\mathcal{SM}^{w_k}_{S^g_c} \leftarrow CE(\mathcal{SM}^{w_k}(\mathcal{S}^g_c), y_c); \nabla \mathcal{M}^{w_k}_{\mathcal{LB}^{w_k}_c} \leftarrow \\CE(\mathcal{SM}^{w_k}(\mathcal{LB}^{w_k}_c), y_c) \\
    l = \sum^{n_c}_{c=1} ML(\nabla \mathcal{M}^{w_k}_{S^g_c}, \nabla \mathcal{M}^{w_k}_{\mathcal{LB}^{w_k}_c}); \ \nabla \mathcal{S}^{g'}_{w_k} = \frac{\partial l}{\partial \mathcal{S}^g} \\
    \mathcal{LB}^{w_k} \leftarrow || \nabla \mathcal{S}^g_{w_k} - \nabla \mathcal{S}^{g'}_{w_k} ||_2
\end{gather}

\noindent
where the matching between $\nabla \mathcal{S}^g_{w_k}$ and $\nabla \mathcal{S}^{g'}_{w_k}$ is performed using the $L_2$ norm $||\cdot||_2$.

To guarantee the preservation of the privacy of the workers from an honest but curious server adoping the method above, we include in our framework an enhanced Local Differential Privacy (LocalDP) strategy.
Traditional LocalDP can fit the considered scenario by adding $Gaussian$ or $Laplacian$ noise and clipping to the obtained updates of $\mathcal{S}^g$. 
For our approach to be effective and, at the same time, to preserve the performance of the distilled data we have to estimate the amount of both {\em(i)} the noise and the {\em(ii)} clipping level.
To achieve this, a worker must perform a grid search over various noise and clipping parameters while simultaneously evaluating performance retention and testing the attack effectiveness under these conditions. However, the computational overhead introduced by this exhaustive search could significantly slow down the distillation process.

To overcome this limitation, we propose an optimized strategy based on a community-oriented Label Differential Privacy (LabelDP) method, inspired by Arazzi et al. \cite{arazzi2025defense}. Our approach, called LabelDP Obfuscation via Randomized Linear Dispersion (LDPO-RLD, hereafter), enables workers to obscure the correlation between distilled and real images. Instead of using one-hot encoded labels to compute gradient matching, workers redistribute an $\epsilon$ fraction of the primary label's probability across a set of $k$ randomly selected labels using a linear function, $Lin(\cdot)$. 
To generate noisy labels, the authors of~\cite{arazzi2025defense} propose employing Knowledge Distillation through a pre-trained teacher network. However, the use of a teacher network would add unnecessary overhead, particularly in resource-constrained environments.
In our scenario, such an additional complexity is not required, as we use the noisy-label approach to just obfuscate the correlation between real and distilled images. Therefore, we can select the $k$ labels at random. This modification reduces the complexity while maintaining the intended obfuscation level. The resulting changes in the loss and gradient calculations are as follows:

\begin{gather}
\nabla\mathcal{SM}^{w_k}_{S^g_c} \leftarrow CE(\mathcal{SM}^{w_k}(\mathcal{S}^g_c), Lin(y_c, k, \epsilon)); \\
\nabla \mathcal{M}^{w_k}_{\mathcal{LB}^{w_k}_c} \leftarrow CE(\mathcal{SM}^{w_k}(\mathcal{LB}^{w_k}_c), Lin(y_c, k, \epsilon))
\end{gather}

With this new formulation, an attacker attempting to reverse the distillation process to recover the original batch of images cannot accurately reconstruct the worker's scenario, as the specific label distribution used remains unknown.

In Section \ref{sec:experiments}, we present experimental results evaluating the modified loss against the reference Data Leakage attack, along with the final performance of the distilled dataset on the test set.

\section{Experiments and Results}
\label{sec:experiments}

In this section, we discuss the experiments carried out to assess the performance of our framework. Specifically, in Section~\ref{sub:experiments-datasets}, we describe the datasets and the metrics used for our experimental campaign. Section~\ref{sub:experiment_performance} is dedicated to evaluating the performance of our solution, and finally, in Section~\ref{sub:expSecurity}, we show the performance of our defense strategy (LDPO-RLD) against both server and client-side attacks.

\subsection{Experimental Setup}
\label{sub:experiments-datasets}
In our experiments, we utilized five state-of-the-art datasets, selected for their diversity in image content and classification complexity, allowing for a comprehensive evaluation of our SFDD approach. Below, we provide details on each of the five datasets used:

\begin{itemize}
    \item \textbf{MNIST} \cite{deng2012mnist} consists of $70,000$ grayscale images of handwritten digits (0-9), with $60,000$ images designated for training and $10,000$ for testing. Each image is 28x28 pixels in size.

    \item \textbf{CIFAR-10} \cite{krizhevsky2009learning} contains $60,000$ 32x32 color images categorized into 10 classes, with $50,000$ images for training and $10,000$ images for testing.

    \item \textbf{SVHN (Street View House Numbers)} \cite{netzer2011reading} is a real-world image dataset designed for digit recognition in natural scene images. The dataset is provided by Stanford University and is divided into $73,257$ training and $26,032$ testing images. Additionally, there are $531,131$ extra images, which are considered somewhat less challenging by the dataset provider; however, these additional images were not used in our experiments.

    \item \textbf{GTSRB (German Traffic Sign Recognition Benchmark)} \cite{stallkamp2012man} contains $39,270$ images of traffic signs categorized into $43$ classes. The dataset is divided into $26,640$ training images and $12,630$ testing images.  

    \item \textbf{CIFAR-100} \cite{krizhevsky2009learning} is similar to CIFAR-10, CIFAR-100 contains $60,000$ 32x32 color images, but with $100$ classes instead of $10$. Each class has $600$ images, divided into $500$ training images and $100$ testing images per class.
\end{itemize}

The following metrics are employed to quantify system performance:

\begin{itemize}
    \item \textbf{Accuracy} is defined as the proportion of correctly predicted instances out of the total number of instances:
    
    $$Accuracy = \frac{TP}{TP+FP}$$
    
    where TP (True Positives) refers to the number of instances where the model correctly predicts the positive class, whereas FP (False Positives) refers to the number of times that the model incorrectly predicts a positive class.
    \item \textbf{Mean Squared Error (MSE)} measures the average squared difference between the estimated values and the true value:

    $$MSE= \frac{1}{n}\sum_{i=1}^n(y_i-\hat{y_i})^2$$
    where $n$ is the number of data points, $y_i$ is the actual (true) value, $\hat{y_i}$ is the predicted (estimated) value, and the squared difference $(y_i-\hat{y_i})^2$ measures the error for each prediction.

    \item \textbf{Attack Success Rate (ASR)} refers to the percentage of times an attack achieves its intended goal or outcome. In the context of backdoor attacks like the Doorping attack, the success rate would be the proportion of instances in which the attack successfully manipulates the model or dataset as intended by the attacker.
    
\end{itemize}
\subsection{Performance Analysis}
\label{sub:experiment_performance}
In this section, we analyze the results of the experiments designed to evaluate the performance of our approach. As outlined in Section~\ref{sub:centralizedDistillation}, the distillation process requires two models, one for distillation and the second for the performance evaluation of the distilled images. In our performance analysis, we employ the same CONVNET architecture described in~\cite{zhao2020dataset} for both models. This architecture features multiple convolutional layers, each followed by normalization, activation, and pooling layers. The design of the CONVNET allows for extensive customization, including the number of filters (\texttt{net\_width}), network depth (\texttt{net\_depth}), activation functions (\texttt{net\_act}), normalization techniques (\texttt{net\_norm}), and pooling strategies (\texttt{net\_pooling}). The input images are standardized to a size of \(32 \times 32\) pixels. We maintained consistent network settings for both the distillation and evaluation models across the centralized version and the federated clients.

In the first experiment, we compared the performance in terms of the accuracy result of the standard DD solution against our SFDD approach with $5$ clients as a baseline.
To fulfill the assessment, we evaluated both solutions by changing the setting for the produced Images Per Class (IPC). In particular, for each considered dataset, we collected the results producing $1$, $10$, and $50$ images for each class.

\begin{table}[!ht]
\centering
\begin{tabular}{||l|c|c|c||}
\hline \hline
\textbf{Dataset} & \textbf{IPC} & \textbf{Centralized} \cite{zhao2020dataset} & \textbf{SFDD} \\ \hline
\multirow{2}{*}{MNIST}    & 1       & $91.92\%$      & 92.02\%    \\
                          & 10      & 97.49\%      & 97.58\%    \\ 
                          & 50      & 98.30\%      & 98.72\%    \\ \hline
\multirow{2}{*}{CIFAR10}  & 1       & 27.82\%      & 28.10\%     \\
                          & 10      & 44.29\%      & 43.65\%    \\ 
                          & 50      & 53.14\%      & 53.40\%    \\ \hline
\multirow{2}{*}{SVHN}     & 1       & 30.58\%      & 29.10\%     \\
                          & 10      & 75.19\%      & 75.89\%    \\ 
                          & 50      & 81.70\%      & 81.58\%    \\ \hline
\multirow{2}{*}{GTSRB}    & 1       & 31.94\%      & 32.13\%    \\
                          & 10      & 66.55\%      & 65.38\%    \\ 
                          & 50      & 75.64\%      & 71.08\%    \\ \hline
\multirow{2}{*}{CIFAR100} & 1       & 12.41\%      & 12.57\%    \\
                          & 10      & 24.82\%      & 24.68\%    \\ 
                          & 50      & 29.73\%      & 29.43\%    \\ \hline \hline
\end{tabular}
\caption{Comparison of the performance of the Centralized DD approach and SFDD}
\label{tab:distillation-performance}
\end{table}

As shown in Table \ref{tab:distillation-performance}, our SFDD approach consistently achieved comparable performance across the five different datasets. For the MNIST dataset, our approach slightly surpasses the centralized approach with 1 and 10 IPC and remains highly competitive with 50 IPC. In the CIFAR10 dataset, SFDD performs slightly better with 1 IPC and maintains very close performance with 10 and 50 IPC. The SVHN dataset results show that SFDD is nearly as effective as the centralized method with 1 IPC and slightly outperforms it with 10 IPC. For the GTSRB dataset, our method exhibits superior performance with 1 IPC and shows strong results with 50 IPC. In the CIFAR100 dataset, the SFDD performs admirably with 1 IPC and is nearly identical in performance with 10 IPC. Overall, our SFDD approach proves to be a robust and effective method, consistently delivering performance on par with the state-of-the-art centralized method.

The second experiment aims to assess whether the number of clients participating in SFDD impacts the quality of the generated images. To evaluate this, we conducted a performance analysis using 10 and 15 clients, alongside the default setting of 5 clients. We measured the final mean accuracy and standard error on test sets across multiple randomly initialized networks trained with the distilled images. For consistency, we set the number of images per class to 10 as the baseline for this experiment.

\begin{figure}[!ht]
    \centering
    \includegraphics[width=0.9\columnwidth]{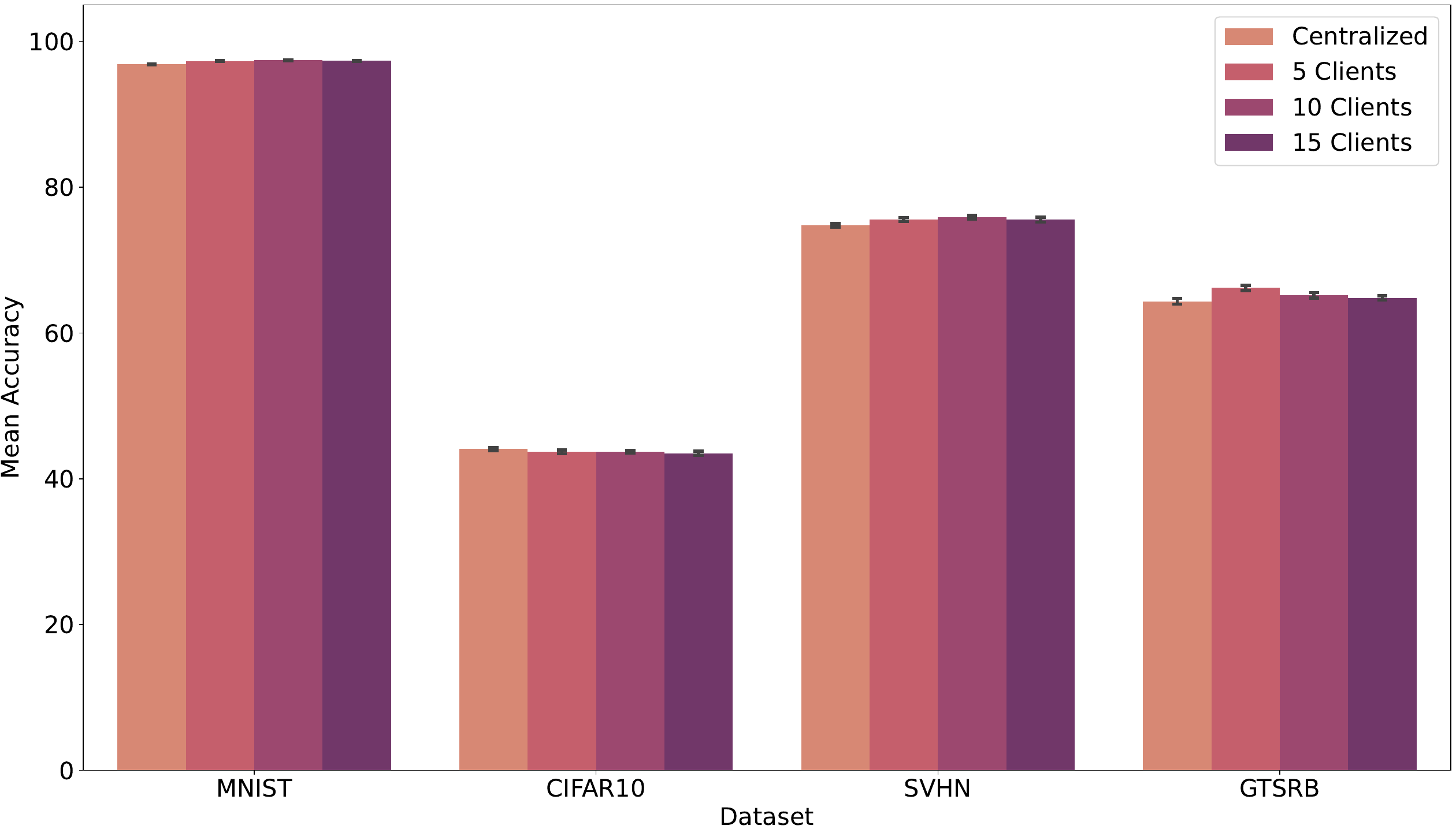}
    \caption{SFDD performance with different numbers of clients}
    \label{fig:clientsAnalysis}
\end{figure}

Figure~\ref{fig:clientsAnalysis} presents the results for this second experiment. Based on this diagram, we can conclude that increasing the number of clients does not significantly impact distillation performance. In MNIST and SHVN, an increase in the number of clients results in slightly better distillation performance. In CIFAR-10 and GTSRB, it results in a slight decrease. Overall, the number of clients does not affect the results, as the differences are negligible.
This finding allows us to guarantee that multiple clients can participate in our framework without affecting the quality of the final results.

\subsection{Security Analysis}
\label{sub:expSecurity}
As for the experiments dealing with the security analysis, we employed the same CONVNET architecture from~\cite{zhao2020dataset} for both distillation and evaluation models, but with certain modifications to simplify the architecture and put the attacker in the best possible scenario for them. Specifically, we discarded the pooling layers and substituted the ReLU activation function with Sigmoid. These modifications are consistent with the changes made in the original paper presenting the Deep Leakage attack \cite{zhu2019deep}. The rationale behind replacing ReLU with Sigmoid is that Sigmoid facilitates better gradient flow, which is advantageous for an attacker. Demonstrating robustness under these conditions suggests that our approach should also be effective against attacks on more complex models.

In our experimental setup, we assume that the attacker (i.e., an honest but curious server) has full knowledge of the initial model parameters used by clients. Additionally, to maintain consistency with the performance analysis, the number of clients in the SFDD approach is set to 5.

The first set of experiments is meant to assess the robustness of the SFDD framework against the Deep Leakage attack \cite{zhu2019deep}. Hence, we conducted preliminary experiments aimed at simulating the most advantageous scenario for an attacker. This involved implementing specific architectural changes in the model that distills the dataset on the client side. As illustrated in Tables \ref{tab:AttackPerformanceMSEs} and \ref{tab:AttackPerformanceDisAccuracy}, our approach empowered with the LDPO-RLD method effectively prevents the attacker from reconstructing the original image used by the client. Indeed, the results clearly show that the Mean Squared Errors (MSEs) increase significantly when using this countermeasure.

\begin{table}[!ht]
   \resizebox{\columnwidth}{!}{
   	 \begin{tabular}{||l|c|c||}
        \hline
        \hline
        \textbf{Dataset} & \textbf{SFDD w/o LDPO-RLD} & \textbf{SFDD with LDPO-RLD} \\
        \hline
        MNIST & 0.245 & 1.575 \\
        \hline
        CIFAR10 &  0.80 &  1.92 \\
        \hline
        SVHN &  0.71 & 2.02 \\
        \hline
        GTSRB &  0.725 & 1.84 \\
        \hline
        CIFAR100 & 0.57 & -  \\
        \hline
        \hline
    \end{tabular}
}
    \centering
    \caption{MSE values between the normalized ground truth batch (from the real dataset) and the normalized reconstructed batch computed during the attack, with and without our defense}
    \label{tab:AttackPerformanceMSEs}
\end{table}

\begin{table}[!ht]
    \begin{tabular}{||l|c||}
        \hline
        \hline
        \textbf{Dataset} & \textbf{Accuracy Variation(\%)}\\
        \hline
        MNIST &  -0.48\% \\
        \hline
        CIFAR10 &  +0.28\% \\
        \hline
        SVHN &  +0.10\% \\
        \hline
        GTSRB & -1.11\% \\
        \hline
        CIFAR100 & +0.27\% \\
        \hline
        \hline
    \end{tabular}
    \centering
    \caption{Accuracy variation of a reference deep learning model trained on the dataset distilled by our approach with the LDPO-RLD defense compared to the same model trained on a dataset distilled using the original centralized distillation scheme}
    \label{tab:AttackPerformanceDisAccuracy}
\end{table}

From these experiments, we can derive two important findings, namely {\em(i)} LDPO-RLD is effective as a countermeasure against deep leakage attacks; and {\em(ii)} LDPO-RLD has a negligible impact on the distillation performance.

The second experiment aims to compare the performance of the standard Local Differential Privacy (LDP) and our LDPO-RLD.

LDP requires each client to perform a grid search to determine the optimal hyperparameter configuration that prevents the attacker from reconstructing the original image. However, this process introduces significant computational overhead, particularly when clients seek a configuration that minimally impacts distillation performance. Finding the best hyperparameters may also require manual intervention, as estimating an appropriate MSE threshold for attack simulations during grid search is challenging (it heavily depends on the dataset's image characteristics). Additionally, to accurately determine an MSE threshold that ensures the desired level of privacy, multiple attack simulations with the same hyperparameters must be conducted to filter out non-significant variations in MSE. In contrast, our LDPO-RLD approach eliminates these overheads entirely. It operates seamlessly without requiring additional settings or manual inspection of grid search results, making it a more efficient, user-friendly, and reproducible solution.

Table \ref{tab:PerformanceDifference} shows the average performance in terms of the variation in the accuracy results between LDP and LDPO-RLD across various datasets. From this table, we can observe that for the MNIST dataset, there is a negligible decrease in performance with LDPO-RLD (-$0.29\%$), which suggests that the two methods perform similarly for simpler datasets. With the CIFAR10 dataset, instead, there is a slight improvement of $0.60\%$, demonstrating that LDPO-RLD marginally outperforms LDP. SVHN experiment has a more significant improvement ($2.56\%$), indicating that our method handles moderately complex datasets better than LDP.
Finally, GTSRB has the highest improvement ($8.94\%$), showing that LDPO-RLD significantly outperforms distillation performance for more complex datasets.

The results show that LDPO-RLD is not only an effective defense against deep leakage attacks but also outperforms LDP in distillation performance. As dataset complexity increases, the performance gap grows, underscoring LDPO-RLD's enhanced robustness and efficiency in real-world applications. These findings demonstrate that our LDPO-RLD countermeasure is both secure and advantageous, improving the overall SFDD framework's performance without the need for additional configurations or increased time complexity.

\begin{table}[!ht]
    \centering
    \begin{tabular}{||c|c||}
        \hline
        \hline
        \textbf{Dataset} & \textbf{Accuracy Variation (\%)} \\
        \hline
        MNIST & -0.29\% \\
        \hline
        CIFAR10 & 0.60\% \\
        \hline
        SVHN & 2.56\% \\
        \hline
        GTSRB & 8.94\% \\
        \hline
        \hline
    \end{tabular}
    \caption{Average performance differences between LDPO-RLD and classical LDP}
    \label{tab:PerformanceDifference}
\end{table}

The final experiment focuses on evaluating the effectiveness of our SFDD strategy against backdoor attacks. Specifically, we tested our solution using the Doorping attack \cite{liu2023backdoor}, which serves as a benchmark for state-of-the-art backdoor attacks targeting dataset distillation. To assess the maximum resilience of our approach, we considered the worst-case scenario where the attacker controls $\frac{z-1}{2}$ of the workers involved. The experiment was conducted across multiple scenarios, varying the number of workers and datasets, to observe how SFDD performs against the Doorping attack under different levels of complexity introduced by the distributed nature of our approach.

\begin{figure}[!ht]
    \centering
    \includegraphics[width=0.9\columnwidth]{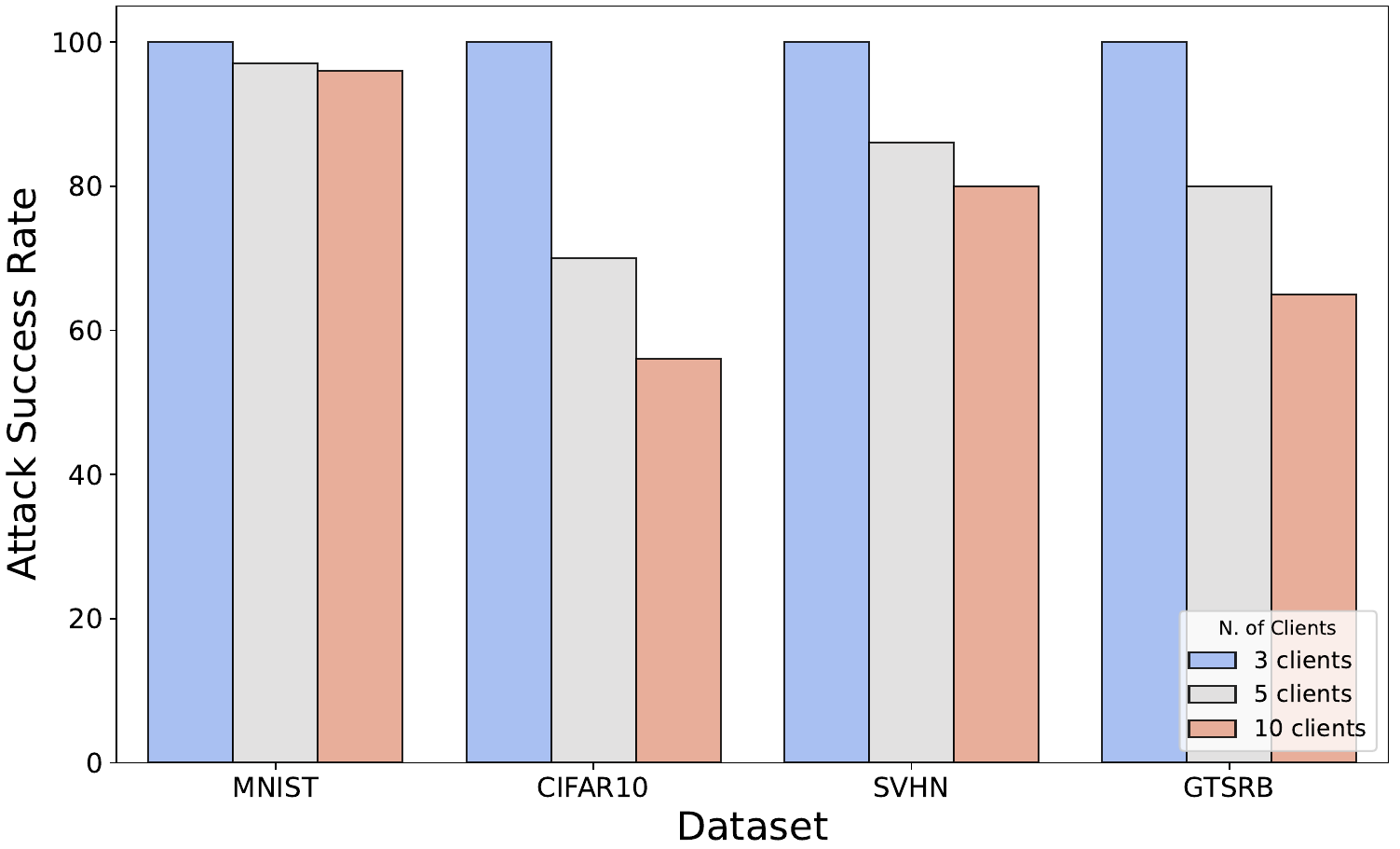}
    \caption{Effect of SFDD on Doorping attack for different datasets}
    \label{fig:experimentOnDoorping}
\end{figure}

In Figure \ref{fig:experimentOnDoorping}, we report the results obtained in a configuration with 3, 5 and 10 clients. In particular, differently from the previous experiment, we report here the results obtained with 3 clients just to show that with a very small number of clients the attack remains effective.
However, such a small number of clients is not common in a Federated Learning scenario, and as the number of clients increases, we observe a general decline in the success rate, particularly with more complex datasets. Remarkably, CIFAR10 and GTSRB exhibit a significant reduction in the attack success rate when the number of clients is the highest considered. These results demonstrate the resilience of our SFDD solution to the Doorping attack, under the assumption of a realistic Federated Learning scenario involving a large number of clients.

\section{Conclusion}
\label{sec:conclusion}

Dataset Distillation (DD) compresses the knowledge of an entire training data into a few synthetic training images. Current architectures for DD require a centralized entity that collects the data to be distilled in a single point of aggregation, thus leading to critical privacy concerns.
To address these risks, we introduce a Secure and Federated Data Distillation (SFDD) framework, inspired by Federated Learning (FL), which decentralizes the distillation process while maintaining privacy. We apply a gradient-matching-based distillation method, modified for a distributed environment where clients participate in the distillation process without sharing raw data. The central aggregator iteratively refines a synthetic dataset by incorporating updates from clients while ensuring data confidentiality. To safeguard against potential inference attacks by the server, which could use gradient updates to reconstruct private data, we integrate an enhanced Local Differential Privacy (LocalDP) approach called LDPO-RLD (LabelDP Obfuscation via Randomized Linear Dispersion). Experiment results demonstrate that the SFDD approach consistently achieved performance
comparable to the classic DD framework across different datasets. Moreover, we prove that the LDPO-RLD is an effective countermeasure against deep leakage attacks and does not affect the distillation performance.
Additionally, we evaluate the framework's resilience to malicious clients carrying out backdoor attacks, such as Doorping, and show our framework robustness in scenarios with a large number of participating clients.

Our solution represents a significant advancement in dataset distillation, allowing diverse data owners to contribute to the construction of a shared synthetic dataset without exposing sensitive information. Future research can explore optimizations in aggregation strategies and further security enhancements to mitigate emerging threats, such as Doorping attacks, in edge scenarios (e.g., scenarios with a very limited number of clients). The SFDD framework paves the way for privacy-aware distributed learning, with applications extending beyond medical data sharing to other fields requiring confidentiality-preserving collaboration.

\section*{Acknowledgments}

This work was supported in part by the project SERICS (PE00000014) under the NRRP MUR program funded by the EU-NGEU. Views and opinions expressed are however those of the authors only and do not necessarily reflect those of the European Union or the Italian MUR. Neither the European Union nor the Italian MUR can be held responsible for them.
\bibliographystyle{plain}

\end{document}